\documentclass[11pt, journal]{IEEEtran}
\usepackage{cite,amsmath,graphicx}
\usepackage{amsmath,amssymb,amsfonts}
\usepackage{algorithmic,algorithm}
\usepackage{graphicx}
\usepackage{textcomp}
\usepackage{xcolor}
\usepackage{arydshln}
\usepackage{multirow}
\usepackage[final]{changes}
\definechangesauthor[name={Zhu}, color=blue]{Zhu}

\def\BibTeX{{\rm B\kern-.05em{\sc i\kern-.025em b}\kern-.08em
    T\kern-.1667em\lower.7ex\hbox{E}\kern-.125emX}}

\begin{document}
\title{\huge 
FreeMark: A Non-Invasive White-Box Watermarking for\\ Deep Neural Networks\\[0.3cm]
}

\author{Yuzhang Chen$^{1,*}$, 
Jiangnan Zhu$^{1,*}$, 
Yujie Gu$^{1}$, 
Minoru Kuribayashi$^{2}$, 
Kouichi Sakurai$^{1}$\\
\thanks{$^{*}$Y. Chen and J. Zhu contributed equally to this work.\\$^{1}$Faculty of Information Science and Electrical Engineering, 
Kyushu University, Fukuoka, Japan\\
$^{2}$Center for Data-driven Science and Artificial Intelligence, Tohoku University, Sendai, Japan}
}

\maketitle

\begin{abstract}
Deep neural networks (DNNs) have achieved significant success in real-world applications. However, safeguarding their intellectual property (IP)  remains extremely challenging. Existing DNN watermarking for IP protection often require modifying DNN models, which reduces model performance and limits their practicality. 

This paper introduces FreeMark, a novel DNN watermarking framework that leverages cryptographic principles \textit{without} altering the original host DNN  model, thereby avoiding any reduction in model performance.
Unlike traditional DNN watermarking methods, FreeMark innovatively generates secret keys from a pre-generated watermark vector and the host model using gradient descent.
These secret keys, used to extract watermark from the model’s activation values, 
are securely stored with a trusted third party, enabling reliable watermark extraction from suspect models. 
Extensive experiments demonstrate that FreeMark effectively resists various watermark removal attacks while maintaining high watermark capacity. \\[-0.42cm]
\end{abstract}




\begin{IEEEkeywords}
deep neural network, copyright protection, white-box, watermarking, non-invasive, intellectual property
\end{IEEEkeywords}


\section{Introduction}

Deep neural network (DNN) watermarking is a typical approach for protecting the intellectual property of DNN models. It allows the model owner to embed a predetermined watermark into the model and later prove ownership by extracting it. Watermarking techniques are broadly divided into two categories: black-box and white-box, depending on whether the model's internal details are available.


Black-box watermarking, which operates without access to the model's internal details, typically relies on backdoor attacks to embed and extract watermarks using specifically designed trigger data, such as pattern-based triggers \cite{r_21}, ODD-based triggers \cite{r_21}, and perturbation-based triggers \cite{r_22}. However, black-box watermarks tend to be less effective against watermark removal attacks \cite{r_13}.


In contrast, white-box watermarking, which has access to the model's internal details, is more robust \cite{r_10}. In this setting, watermarks are embedded into host DNN models using techniques such as parameter regularization \cite{r_9}, adversarial learning \cite{r_11}, residuals of parameters \cite{r_12}, probability density functions \cite{r_13}, hidden memory states \cite{r_14}, and directly modifying the model's weights \cite{r_27}.


However, all existing DNN watermarking methods require modifying the host models, which significantly reduces model performance and limits their practicality. Although efforts have been made to minimize this impact, significant limitations still remain in their real-world applications.
Naturally, this raises the question:
{\textit{Is it possible to watermark DNNs without any decline in model accuracy?}

In this paper, we provide an affirmative answer, demonstrating that DNNs can be watermarked without altering the models. Specifically, we propose a novel watermarking framework, termed FreeMark, which leverages cryptographic principles without modifying the original models, thereby avoiding any reduction in model performance (see Fig.~\ref{fig:Overview}). Moreover, our method FreeMark offers three key advantages.


(i) Non-invasive. FreeMark ingeniously integrates watermark with the intrinsic features of the host model into secret keys stored on a trusted third party (TTP), without making any modifications to the model, thereby eliminating the risk of performance degradation during watermark embedding.

(ii) Robustness and security. Our FreeMark demonstrates strong resistance to watermark removal attacks, such as pruning and fine-tuning. Even under extreme scenarios, FreeMark accurately distinguishes the host model from suspect models.

(iii) Efficiency. FreeMark is computationally simple, offering a lightweight and convenient framework for watermark embedding, extraction, and verification.


\deleted{The main pipeline of our watermarking mechanism is detailed in Fig.~\ref{fig:Overview}. For a well-developed watermark, it firstly needs to correctly insert and extract the predetermined watermark from the host model $H$. Then, the watermark should not affect the model during insertion, and it should still be extracted after certain modifications to the model. Beyond these principles, we propose higher standards: watermark mechanism should ensure only the model owner can correctly extract the predetermined watermark $\boldsymbol{b}$, and the extraction process should be simple and reliable. Additionally, we require that the watermark need to be undetectable to outsiders of the host model after insertion and should not be extracted from unmarked models $H'$, indicating low false positive. Finally, while meeting these requirements, the watermark should have a capacity $N$ that as high as possible.}

\begin{figure*}[!htbp]
\centerline{\includegraphics[width=1.0\textwidth]{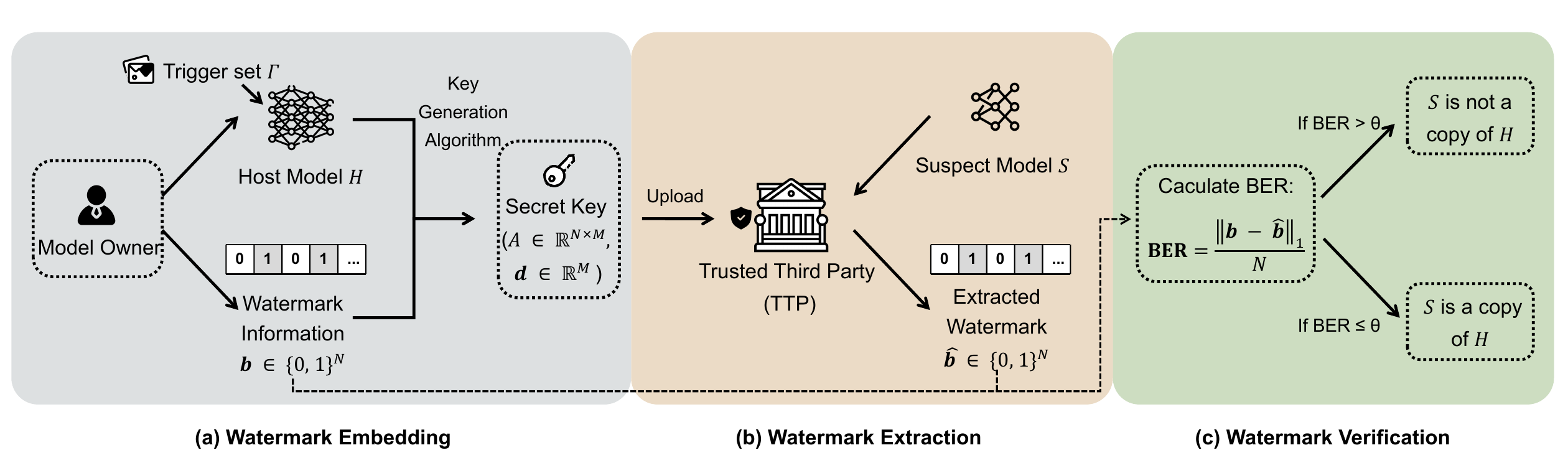}}
\caption{The workflow of our proposed method FreeMark. In watermark embedding, a secret key pair $(A, \boldsymbol{d})$ is generated by integrating the host model $H$ and the watermark $\boldsymbol{b}$. In watermark extraction, TTP employs secret keys to extract watermark $\boldsymbol{\hat{b}}$ from a suspect model $S$. In watermark verification, the BER between original watermark $\boldsymbol{b}$ and the extracted watermark $\boldsymbol{\hat{b}}$ is calculated to determine  whether $S$ is a copy of $H$.}
\label{fig:Overview}
\end{figure*}

\section{Preliminaries}

The following notations will be used throughout this paper.
\begin{itemize}
    \item  $\boldsymbol{b}$: a watermark vector\footnote{All the vectors of length $N$ is of size $N\times 1$ in this paper.} of length $N$, i.e., $\boldsymbol{b} \in \{0,1\}^N$.\\[-0.4cm]
    \item $\boldsymbol{\hat{b}}$: an extracted watermark vector, also $\boldsymbol{\hat{b}} \in \{0,1\}^N$.\\[-0.4cm]
    \item $\Gamma$: a trigger set for watermarking, which 
    is a subset of the training dataset.\\[-0.4cm]
    \item $\boldsymbol{\overline{f}}_l$: 
    the average activation value vector over the trigger set $\Gamma$ in the $l$-th layer with $M$ neurons, i.e., $\boldsymbol{\overline{f}}_l\in \mathbb{R}^M$.\\[-0.4cm]
    \item $(\boldsymbol{A}, \boldsymbol{d}):$ a pair of secret keys, where $\boldsymbol{A}\in \mathbb{R}^{N\times M}$ and $\boldsymbol{d}\in \mathbb{R}^M$. \\[-0.4cm]
    \item $\theta$: a predetermined threshold, where $\theta \in [0, 1]$.
\end{itemize}

\section{Proposed Method}
\label{sec:propose_method}
Our FreeMark method consists of three components: watermark embedding, watermark extraction, and watermark verification, as described below.

\subsection{Watermark Embedding}

\textbf{Preparation.} 
For a pretrained host model $H$, the model owner selects a trigger set $\Gamma$ that contains at least one sample from each label in the training data. By using the trigger set $\Gamma$ as input to the host model $H$, the owner can obtain the average activation value vector $\boldsymbol{\bar{f}}_l$. The model owner also generates a binary watermark vector $\boldsymbol{b}$ of length $N$.  

Define a non-linear function $\Delta(\cdot)$ as
\begin{align*}
    \Delta(x)\triangleq \text{Thresholding}\left(\frac{1}{1 + e^{-x}}\right)
\end{align*}
where 
\begin{align*}
    \text{Thresholding}(z) = 
\begin{cases} 
0 & \text{if } z < 0.5 \\
1 & \text{if } z \geq 0.5.
\end{cases}
\end{align*}
For a vector $\boldsymbol{x}=(x_1,\ldots,x_n)$, denote 
\begin{align*}  \Delta(\boldsymbol{x})=\left(\Delta(x_1),\ldots,\Delta(x_n)\right).
\end{align*}

\begin{algorithm}[tb]
     \caption{Generate Secret Key $\boldsymbol{A}$ via Gradient Descent}
    \label{alg:generateA}
    \textbf{Input}: Watermark vector $\boldsymbol{b}$; an auxiliary vector $\boldsymbol{\mu}$; learning rate $\lambda$; iteration times $T$\deleted{; Current iteration times $t$}. 
    
    \textbf{Output} a secret matrix $\boldsymbol{A}$
    
    \begin{algorithmic}[1] 
        \STATE Initialize $\boldsymbol{A}^{(0)}$ with random values and $t=0$.
        \WHILE{$t<T$}
        \STATE Calculate the current loss 
        $\mathcal{L}_t=\Delta(\boldsymbol{A}^{(t)}\cdot \boldsymbol{\mu})-\boldsymbol{b}.$
        \STATE
        Update $\boldsymbol{A}^{(t)}$ by  $\boldsymbol{A}^{(t+1)}=\boldsymbol{A}^{(t)}-\lambda \cdot \frac{\partial \mathcal{L}_t}{\partial \boldsymbol{A}^{(t)}}$.

        \ENDWHILE
        \STATE \textbf{return} $\boldsymbol{A}$ = $\boldsymbol{A}^{(T)}$.
    \end{algorithmic}
\end{algorithm}

\noindent
\textbf{Generate secret keys.} 
In \cite{r_9}, DeepSigns used a predetermined watermark vector $\boldsymbol{b}$ and a secret matrix $\boldsymbol{A}$ to modify the host model $H$ such that
\begin{align*}
    \boldsymbol{b}\approx \Delta(\boldsymbol{A}\cdot \boldsymbol{f}'_l)
\end{align*}
via fine-tuning, by adding an additional term that minimizes the distance between $\boldsymbol{b}$ and $\Delta(\boldsymbol{A} \cdot \boldsymbol{f}'_l)$ to the overall loss function. Here $\boldsymbol{f}'_l$ is the activation value vector in the $l$-th layer of the modified model $H'$. 
Similarly, other existing watermarking methods also require modifying the host model, e.g. \cite{r_11, r_12,r_13, r_14, r_27}. However, all of these DNN watermarking methods that involve modifying the host model reduce the accuracy and limit their applications.

Our motivation is to design a DNN watermarking mechanism that does not require modifying the host models, thereby preserving model accuracy. 

To that end, we first randomly generate an auxiliary vector $\boldsymbol{\mu} \sim \mathcal{N}(\boldsymbol{0},\boldsymbol{1})$ of length $M$ from the standard normal distribution.
Next, by using the watermark $\boldsymbol{b}$ and the auxiliary vector $\boldsymbol{\mu}$, we aim to generate a secret key $\boldsymbol{A}\in \mathbb{R}^{N\times M}$ such that 
\begin{align*}
    \boldsymbol{b}=\Delta(\boldsymbol{A}\cdot \boldsymbol{\mu}).
\end{align*}
To achieve this, we introduce Algorithm \ref{alg:generateA} via gradient descent method, which can effectively address the non-linear multivariable optimization problem of deriving $\boldsymbol{A}$.
Then we establish another secret key $\boldsymbol{d}\in \mathbb{R}^M$ by 
\begin{align*}
    \boldsymbol{d} = \alpha\boldsymbol{\bar{f}}_l - \boldsymbol{\mu}
\end{align*}
where $\alpha\in \mathbb{R}$ is a scaling factor such that 
\begin{align}\label{eq-condition-for-d}
    \lVert \Delta(\boldsymbol{A}\cdot \boldsymbol{d})-\boldsymbol{b}\rVert_1\ge \theta N
\end{align}
to ensure the security, integrity, and robustness of the watermark.



The overall secret key generation pipeline is illustrated in Algorithm \ref{alg:wm_embedding}. The generated secret key pair $(\boldsymbol{A}, \boldsymbol{d})$ such that
\begin{align*}
    \boldsymbol{b} = \Delta\left(\boldsymbol{A}\cdot (\alpha\boldsymbol{\bar{f}}_l-\boldsymbol{d})\right)
\end{align*}
is uploaded to a TTP, which ensures the secure storage of the secret keys, prevents unauthorized access, and provides a trusted environment for watermark extraction from a suspect model in the later process.

\begin{algorithm}[tb]
    \caption{Key Generation Algorithm}
    \label{alg:wm_embedding}
    \textbf{Input}: Host model $H$; trigger set $\Gamma$; target layer $l$; an auxiliary vector $\boldsymbol{\mu}$; threshold $\theta$.
    
    \textbf{Output} Secret key ($\boldsymbol{A}$, $\boldsymbol{d}$) and scaling factor $\alpha$ 
    
    \begin{algorithmic}[1] 
        \STATE 
        Calculate the average activation value $\boldsymbol{\overline{f}}_l$ over the trigger set $\Gamma$ in the $l$-th layer of the host model $H$.
        \STATE
        Generate secret key $\boldsymbol{A}$ via Algorithm \ref{alg:generateA}.
         \STATE Calculate secret key $\boldsymbol{d} =  \alpha\boldsymbol{\bar{f}}_l - \boldsymbol{\mu}$, where $\alpha$ satisfies \eqref{eq-condition-for-d}.
        \STATE \textbf{return} $(\boldsymbol{A}, \boldsymbol{d})$ and $\alpha$.
    \end{algorithmic}
\end{algorithm}

\subsection{Watermark Extraction}
Algorithm \ref{alg:wm_ex} demonstrates the watermark extraction procedure from a suspect model, which involves two primary steps conducted by the TTP. 

First, by inputting the trigger set $\Gamma$ into the suspect model $S$, 
the average activation value vector $\boldsymbol{\hat{f}}_l$ over $\Gamma$ in the $l$-th layer of $S$ is obtained. 
Second, by using the secret key pair $(\boldsymbol{A}, \boldsymbol{d})$, the TTP computes the extracted watermark information
$\boldsymbol{\hat{b}}=\Delta(\boldsymbol{A}\cdot (\alpha\boldsymbol{\hat{f}}_l - \boldsymbol{d}))$.

\begin{algorithm}[tb]
     \caption{Watermark Extraction}
    \label{alg:wm_ex}
    \textbf{Input}: Suspect model $S$; trigger set $\Gamma$; secret key pair $(\boldsymbol{A}, \boldsymbol{d})$; target layer $l$; scaling factor $\alpha$.
    
    \textbf{Output} Extracted watermark $\boldsymbol{\hat{b}}$ 
    
    \begin{algorithmic}[1] 
        \STATE 
        Calculate the average activation value $\boldsymbol{\hat{f}}_l$ over the trigger set $\Gamma$ in the $l$-th layer of the suspect model $S$. 
        \STATE
        Calculate  $\boldsymbol{\hat{b}}=\Delta(\boldsymbol{A}\cdot (\alpha\boldsymbol{\hat{f}}_l - \boldsymbol{d})).$ 
        \STATE \textbf{return} the extracted watermark $\boldsymbol{\hat{b}}$.
    \end{algorithmic}
\end{algorithm}

\subsection{Watermark Verification}

To verify whether the suspect model $S$ is a copy of the host model $H$, we compare the predetermined watermark $\boldsymbol{b}$ with the extracted watermark $\boldsymbol{\hat{b}}$ from $S$ using the metric of bit-error-rate (BER), where
\begin{align}
    \text{BER} \triangleq \frac{\lVert 
 \boldsymbol{b}- \boldsymbol{\hat{b}} \rVert_1}{N}.
\end{align}
We use a predetermined threshold $\theta \in [0, 1]$ as a criterion. If $\text{BER} \leq \theta$, the suspect model $S$ is regarded as a copy of the host model $H$. Conversely, if $\text{BER} > \theta$, $S$ is not regarded as a copy of $H$. 


\section{Experiments And Results}

\textbf{Setting.} We conduct experiments using MNIST\cite{r_32}, CIFAR-10, and CIFAR-100\cite{r_33} datasets for image classification. For the host model, we use Lenet-5\cite{r_28}, VGG16\cite{r_29}, Resnet20\cite{r_30}, CNN\footnote{Sharing the same structure as the CNN used in \cite{r_9}.}, and WRN-28-10\cite{r_31}.

We set the watermark length to $N=512$, which indicates a significantly larger watermark capacity than existing white-box DNN watermarking schemes\cite{r_9,r_13}. 


Since our method FreeMark does not modify the host model itself, there is no decline in model accuracy during watermarking. Instead, we focus on verifying the following three aspects of the watermarking scheme. 

\textbf{Security.} The watermark $\boldsymbol{b}$ can be accurately extracted using the secret key $(\boldsymbol{A}, \boldsymbol{d})$, while a third party without the secret key cannot accurately extract the watermark.

\textbf{Integrity.} The watermark cannot be extracted from a third party's  unmarked model, indicating that our FreeMark method has a low false-positive rate.

\textbf{Robustness.} The watermark can still be correctly extracted even after the host model undergoes watermark removal attacks such as fine-tuning and model pruning.

 
\begin{table*}[!htbp]
\caption{BER results. ($\uparrow$ indicates that a higher value is better, while $\downarrow$ indicates that a lower value is preferred.) }
\begin{center}
\begin{tabular}{c|c|c|c|c|c|c}
\hline
&Models&Lenet5&VGG16&Resnet20 & CNN&WRN\\ \hline 
\multirow{3}{*}{Security}&Dataset& MNIST& CIFAR-10& CIFAR-100& CIFAR-10&CIFAR-10\\ 
&Accuracy  & 0.9835& 0.8252&   0.5792& 0.7676&0.8470\\ 
&BER with correct secret keys $\downarrow$& 0.0& 0.0& 0.0& 0.0&0.0\\ \hline
\multirow{2}{*}{Integrity}&BER for different hyperparameters $\uparrow$& 0.4805& 0.4941& 0.4785& 0.5039&0.4843\\ 
&BER for different training data $\uparrow$& 0.4746& 0.5059&   0.5215& 0.4844&0.5156\\ \hline
\multirow{4}{*}{Robustness}&Accuracy after model fine-tuning& 0.9862& 0.8599&   0.5966& 0.7703&0.8178\\ 
&BER after model fine-tuning $\downarrow$& 0.0& 0.0& 0.0& 0.0&0.0\\ \cline{2-7}
&Accuracy after pruning by $\eta=0.1$& 0.9789& 0.8215&   0.5789& 0.6636&0.8474\\ 
&BER after model pruning by $\eta=0.1$ $\downarrow$& 0.0& 0.0& 0.0& 0.0&0.0\\ \hline

\end{tabular}
\label{tab:ber}
\end{center}
\end{table*}

\subsection{Experiments on Security}
\textbf{Watermark extraction with correct keys. } 
We conduct experiments to verify that the watermark $\boldsymbol{b}$ can indeed be extracted from the host model $H$ using the correct secret key pair $(\boldsymbol{A}, \boldsymbol{d})$. As shown in Table \ref{tab:ber}, the BER in this scenario is consistently $0$, indicating the security of  FreeMark.

 

\textbf{Watermark extraction with forged keys.} 
We verify that the predetermined watermark cannot be correctly extracted from the host model $H$ using forged keys. To do this, we generate forged keys $\boldsymbol{A}'$ and $\boldsymbol{d}'$ using a standard normal distribution. Additionally, we randomly generate $200$ forged key pairs for each model to conduct the experiments. The BER results in this scenario are expected to be as high as possible.

\begin{figure}[!htbp]
\centerline{\includegraphics[width=0.42\textwidth]{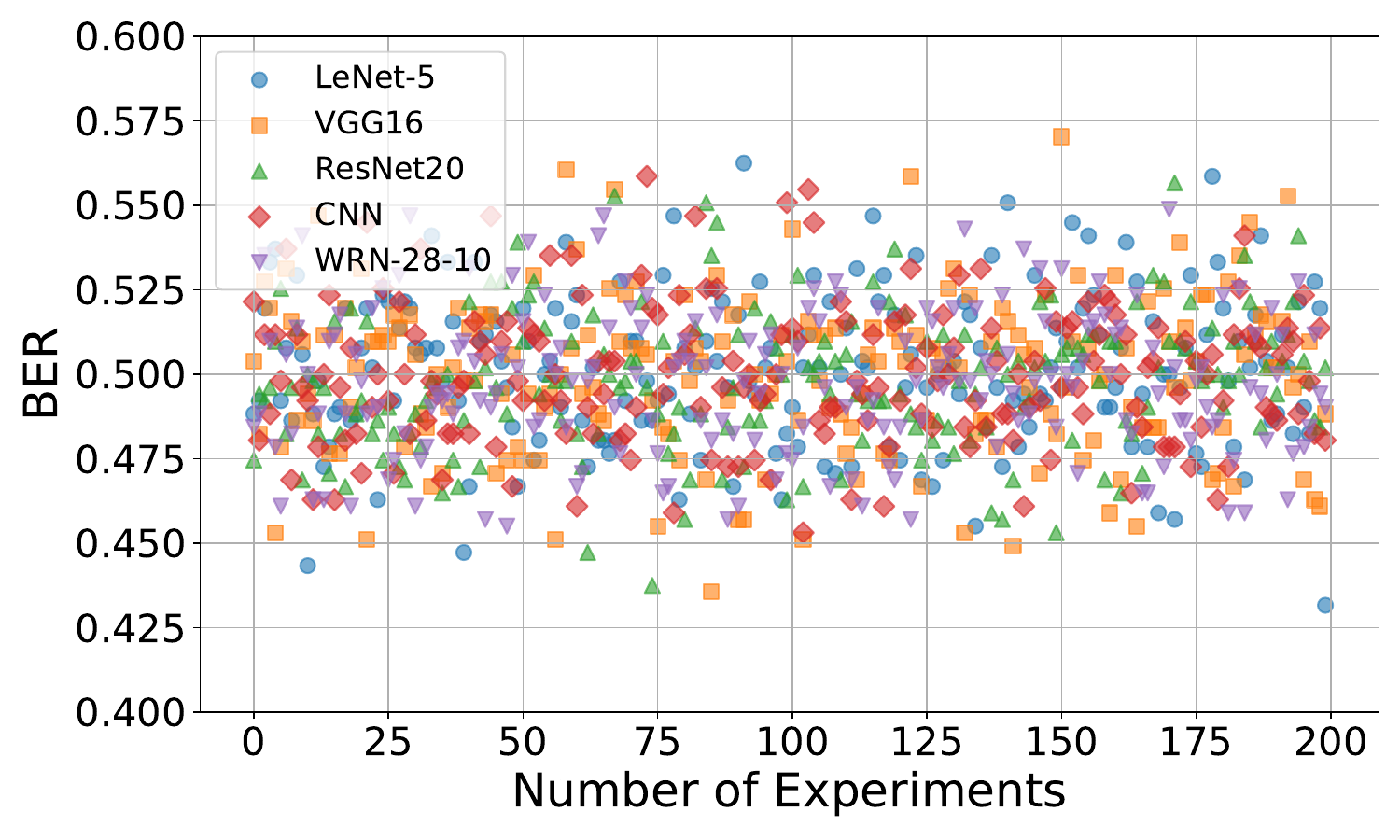}}
\caption{BER with forged secret keys.}
\label{fig:kf}
\end{figure}

Indeed, as shown in Fig. \ref{fig:kf}, the BER for extracting watermarks from all models ranges between $0.4$ and $0.6$, indicating that the extracted watermark is nearly random, thus demonstrating our watermarking method FreeMark ensures 
accurate watermark extraction only with the correct keys.



\subsection{Experiments on Integrity}
We verify that in FreeMark, watermark information will not be erroneously extracted from a third party's model. To strengthen this argument, we conduct experiments in the following two most extreme scenarios:

(i) The third party's model has the same architecture, the same tasks, and uses the same training data as the host model, but is trained with different hyperparameters.
    
(ii) The third party's model has the same structure, the same tasks, and uses the same training hyperparameters as the host model, but is trained on different data.

As shown in Table \ref{tab:ber}, even in the most extreme scenarios, our method FreeMark can accurately distinguish between the host model and a third party's unmarked model, demonstrating that FreeMark has a low false-positive rate.

\subsection{Experiments on Robustness }
To verify the robustness of FreeMark, we conduct experiments on watermark extraction from the host model subjected to three typical watermark removal attacks, as described below. In this scenario, even if the host model is attacked, the watermark is expected to be correctly extracted.


\textbf{Fine-tuning.} 
For each model, we fix a portion of the model's weights and fine-tune the remaining weights using the original training data. As shown in Table \ref{tab:ber}, fine-tuning increases the accuracy of all models except for the WRN. Moreover, we can still accurately extract the watermark from the models that underwent fine-tuning.

 
\textbf{Pruning.} 
A model pruning attack removes all weights below a threshold $\eta$ from the model. As shown in Table \ref{tab:ber} with $\eta=0.1$, the model accuracy decreases, but the watermark can still be correctly extracted.

 

Furthermore, Fig. 3 illustrates the model performance for different pruning thresholds. The results indicate that even after substantial performance degradation due to pruning, the watermark can still be accurately extracted with $\text{BER}=0$.

\begin{figure}[!htbp]
\centerline{\includegraphics[width=0.47\textwidth]{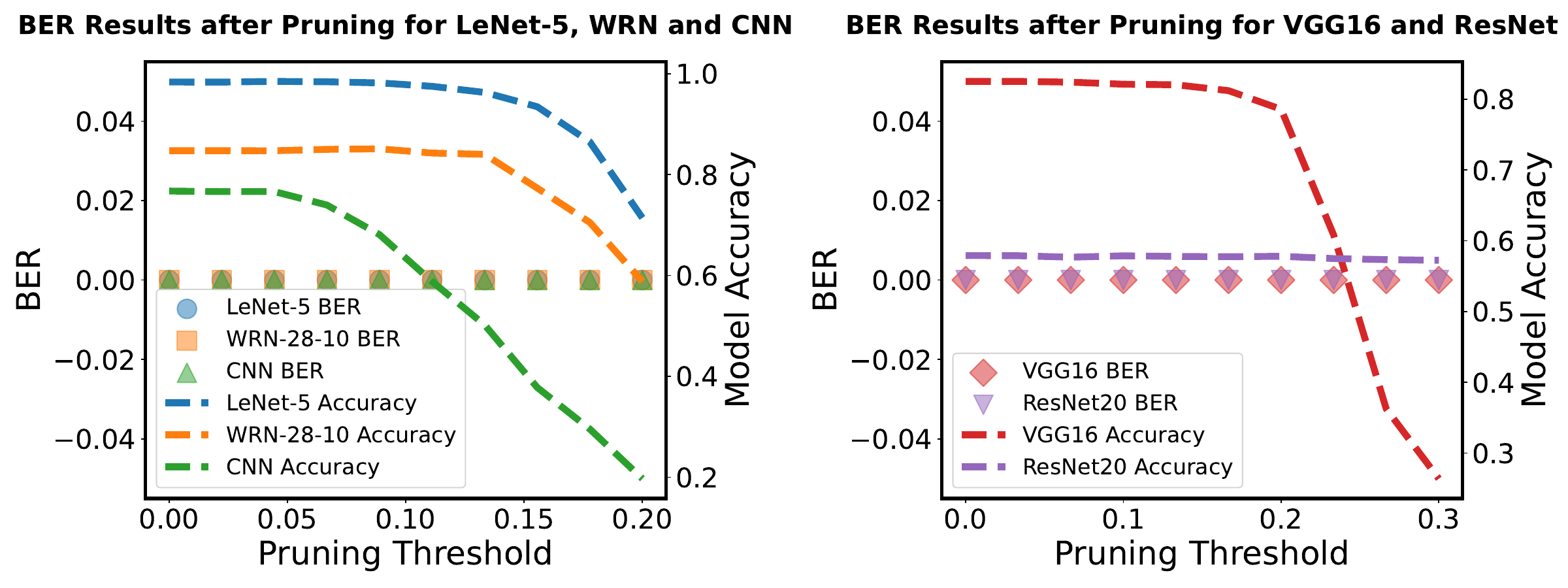}}
\caption{BER after pruning.}
\label{fig3}
\end{figure}

\textbf{Overwriting.} 
A watermark overwriting attack adds multiple watermarks to the model, thereby overriding the original watermark and making it unextractable \cite{r_18}. However, our method FreeMark is completely immune to this type of attack. 
FreeMark does not modify the model itself; instead, it transfers the integration of the watermark and host model into secret keys, which are stored on a TTP. As a result, the watermark cannot be overwritten. Furthermore, FreeMark allows embedding a sufficient number of watermarks into the model, demonstrating the flexibility of watermarking with FreeMark.


\section{Conclusion}
In this paper, we proposed a novel DNN watermarking approach, termed FreeMark. It ingeniously integrates the predetermined watermark and model characteristics into secret keys stored on a TTP, without modifying the model, thereby preserving its original accuracy. Comprehensive experiments demonstrate that FreeMark exhibits superiority in security, integrity, and robustness against various attacks.   This work offers new insights into DNN watermarking and opens up possibilities for watermarking DNNs without model modification.

\end{document}